\documentstyle[preprint,aps]{revtex}
\draft
\begin{document}
\title{Exact solutions of two-band models of graded-gap superlattices}
\author{B.\ M\'{e}ndez and F. Dom\'{\i}nguez-Adame}
\address{Departamento de F\'{\i}sica de Materiales,
Universidad Complutense,
28040 Madrid, Spain}

\maketitle

\begin{abstract}

We have theoretically investigated two-band models of graded-gap
superlattices within the envelope-function approximation. Assuming that
the gap varies linearly with spatial coordinate, we are able to find
exact solutions of the corresponding Dirac-like equation describing the
conduction- and valence-band envelope-functions. The dispersion relation
inside allowed miniband of the superlattice may be expressed in terms of
confluent hypergeometric functions in a closed form.

\end{abstract}


\narrowtext

During the last years, graded-gap superlattices have been the subject of
very detailed investigations as interesting materials for device
applications \cite{Capasso,Brum}.  The graded doping creates a
modulation of both conduction- and valence-bands, which may be
approximated by a sawtooth potential.  The miniband structure can be
obtained within the envelope-function approximation \cite{Bastard}, the
system being usually described by a scalar Hamiltonian
(Schr\"odinger-like) corresponding to decoupled bands.  However, this
approach cannot adequately describe those graded-gap superlattices whose
band modulation is comparable to the magnitude of the gap, and a more
realistic band structure is essential to properly describe the
electronic structure.  In this paper we calculate the miniband structure
of graded-gap superlattices within a two-band model, which is known to
be valid in a large variety of semiconductor superlattices where the
coupling of bands is not negligible, as occurs in some narrow-gap III-V
compounds (InAs, InSb, GaSb).

We obtain the miniband structure in the superlattice by means of the
effective-mass $\bf k\cdot p$ approximation.  There are two coupled
envelope-functions describing the conduction-band and valence-band
states of the semiconductor, subject to an effective $2\times 2$
Dirac-like equation along the growth direction $z$
\begin{equation}
\left( -i\hbar v \sigma_x\partial +{1\over 2}E_g(z)\sigma_z-E \right)
\left( \begin{array}{c} f_c (z)\\ f_v(z)\end{array} \right) =0,
\label{Dirac}
\end{equation}
where $\partial =d/dz$, $\sigma_x$ and $\sigma_z$ are Pauli matrices,
and $E_g(z)$ stands for the position-dependent gap in the two-band
semiconductor superlattice.  The spatial periodicity of the lattice
implies that $E_g(z+L)=E_g(z)$, $L$ being the period of the
superlattice.  We assume that the centre of the gap remains unchanged
when doping; this simplifies calculations and is a good approximation in
several cases (for instance, in GaAs-Ga$_{1-x}$Al$_x$As the centre of
the gap varies only 10\% of the gap difference in both materials).  The
velocity $v=(E_g/2m^*)^{1/2}$ is almost constant in direct gap III-V
semiconductors, and we will assume this constancy hereafter.  In
graded-gap structures the gap varies linearly with position so that
we can write
\begin{equation}
E_g(z)=E_{g0}+(E_{gL}-E_{g0})\left( \frac{z}{L} \right)
\equiv E_{g0}+\Delta E_g \left( \frac{z}{L} \right),
\label{gap}
\end{equation}
where $E_{g0}=E_g(0)$ and $E_{gL}=E_g(L)$ for $0<z<L$.  Note that
$E_g(z)$ is equivalent to a relativistic scalar-like potential in the
Dirac theory, that is to say, Eq.~(\ref{Dirac}) is analogous to the
Dirac equation for a relativistic particle with a position-dependent
mass.  We exploit this analogy to find the exact solutions of
Eq.~(\ref{Dirac}) for the graded-gap (\ref{gap}).  It is well-known that
the Dirac equation for linear scalar-like potentials admits exact
solutions (see Refs.~\onlinecite{yo1,yo2} and references therein) and we
can use a similar method to solve Eq.~(\ref{Dirac}).  Therefore we
define
\begin{equation}
\left( \begin{array}{c} f_c (z)\\ f_v(z)\end{array} \right) =
\left( -i\hbar v \sigma_x\partial +{1\over 2}E_g(z)\sigma_z+E \right)
\left( \begin{array}{c} i\phi (z)\\ \phi(z)\end{array} \right),
\label{phi}
\end{equation}
and inserting (\ref{phi}) in (\ref{Dirac}) one obtains the equation
for the function $\phi (z)$ as
\begin{equation}
\left[ -\,\hbar^2 v^2 \partial^2 + {1\over 4} E^2_g(z) -E^2
-\, \left( {\hbar v \Delta E_g \over 2L} \right)\right] \phi(z)=0.
\label{kg}
\end{equation}

The equation (\ref{kg}) may be reduced to a standard form, the equation
of the parabolic cylinder, by making the change of parameters according
to
\begin{mathletters}
\label{change}
\begin{eqnarray}
x&=&\sqrt{L\over \hbar v \Delta E_g}\, \left( E_{g0}+
\Delta E_g {z\over L} \right), \label{changea}\\
\eta &=& \frac{E^2L}{\hbar v \Delta E_g}. \label{changeb}
\end{eqnarray}
\end{mathletters}
On making these substitutions one gets
\begin{equation}
\frac{d^2\phi (x)}{d\,x^2} + \left( -\,{x^2\over 4}+\eta+{1\over 2}
\right) \phi (x)=0,
\label{parabol}
\end{equation}
whose two independent solutions are parabolic cylinder functions $D_\eta
(x)$ and $D_\eta(-x)$. Using Eq.~(\ref{phi}) we find that the
envelope-functions in the conduction- and valence-bands can be cast in
the matrix form
\begin{equation}
\left( \begin{array}{c} f_c (z)\\ f_v(z)\end{array} \right) =
{\bf D}[x(z)] \left( \begin{array}{c} A\\ B\end{array} \right),
\hspace{1cm}  0<z<L,
\label{result}
\end{equation}
where $A$ and $B$ are arbitrary constants and the $2\times 2$ matrix
${\bf D}[x(z)]$ is written out explicitly as
\begin{equation}
{\bf D}[x(z)]= \left[ \begin{array}{cc}
-i\{ D_\eta(x)-\sqrt{\eta} D_{\eta-1}(x) \}&
-i\{ D_\eta(-x)+\sqrt{\eta} D_{\eta-1}(-x) \}\\
D_\eta(x)+\sqrt{\eta} D_{\eta-1}(x) &
D_\eta(-x)-\sqrt{\eta} D_{\eta-1}(-x) \end{array} \right].
\label{matrix}
\end{equation}

Once the general solution of the Dirac equation (\ref{Dirac}) is
obtained, appropriate boundary conditions should be used to find
eigenenergies. We assume the continuity of the envelope-functions at the
interface $z=L$, namely,
\begin{equation}
\left( \begin{array}{c} f_c (L^-)\\ f_v(L^-)\end{array} \right) =
\left( \begin{array}{c} f_c (L^+)\\ f_v(L^+)\end{array} \right),
\label{conti}
\end{equation}
along with the Bloch condition in the growth direction
\begin{equation}
\left( \begin{array}{c} f_c (L)\\ f_v(L)\end{array} \right) =
\exp(ikL)\, \left( \begin{array}{c} f_c(0)\\ f_v(0)\end{array} \right),
\label{Bloch}
\end{equation}
where $k$ denotes the component of the momentum along the growth
direction $z$. By means of the general solution (\ref{result}) we can
find the dispersion relation as
\begin{equation}
\cos kL = {1\over 2} \mbox{Tr} \left( {\bf D}^{-1}(x_L) {\bf D}(x_0)
\right) \label{tr}
\end{equation}
where for brevity we have defined
\begin{equation}
x_0 = \sqrt{E^2_{g0}L \over \hbar v \Delta E_g}, \hspace{1cm}
x_L = \sqrt{E^2_{gL}L \over \hbar v \Delta E_g}.
\label{x}
\end{equation}
Finally, using the relationship between parabolic cylinder functions and
the confluent hypergeometric functions $M(\alpha,\beta;t)$ \cite{Abra},
it is straightforward although somewhat tedious to demonstrate that the
dispersion relation can be expressed as
\begin{eqnarray}
\cos kL&=& {e^{-(x_0^2+x_L^2)/4}\over 2}
\left\{
M(-\, {\eta \over 2}, {1\over 2}; {x_0^2\over 2})
M({1-\eta \over 2}, {1\over 2}; {x_L^2\over 2})
+ M(-\, {\eta \over 2}, {1\over 2}; {x_L^2\over 2})
M({1-\eta \over 2}, {1\over 2}; {x_0^2\over 2}) \right. \nonumber \\
+\eta x_0 x_L &&\hspace{-5mm}  \left. \left[
M({1-\eta \over 2}, {3\over 2}; {x_0^2\over 2})
M(1-{\eta \over 2}, {3\over 2}; {x_L^2\over 2})+
M({1-\eta \over 2}, {3\over 2}; {x_L^2\over 2})
M(1-{\eta \over 2}, {3\over 2}; {x_0^2\over 2})\right] \right\}.
\label{porfin}
\end{eqnarray}
Whenever the absolute value of the right-hand-side of this equation
is less than unity, a real value of $k$ is found and hence the
dispersion relation inside allowed minibands is obtained. Conversely, if
the absolute value is larger than unity, the energy corresponds to a
minigap of the superlattice.

As a specific example we have considered graded structures with
$E_{g0}=0.18\,$eV (corresponding to InSb), $E_{gL}=0.27\,$eV, $\hbar
v=0.70\,$eV\,nm and superlattice periods $L$ ranging from $5\,$nm up to
$40\,$nm. Results of the allowed minibands and minigaps as a function
of the lattice period are shown in Fig.~\ref{fig1}. Note that allowed
minibands shrink on increasing superlattice period due to the reduction
of the overlap of neighbouring cells.

In conclusion, we have described theoretically the miniband structure in
graded-gap superlattices within a two-band semiconductor model, that is,
we have taking into account the coupling between the conduction- and
valence-bands in the host semiconductor. Assuming that the gap increases
linearly with position, we are able to solve exactly the $2\times 2$
Dirac-like equation of the model. The dispersion relation inside allowed
minibands may be expressed in a closed form in terms of the confluent
hypergeometric functions.

\begin{figure}
\caption{Miniband structure of a graded-gap superlattice within the
two-band model, as a function of the superlattice period. Energies are
measured from the bottom of the conduction band at $z=0$.
Shaded areas correspond to allowed minibands in the superlattice.}
\label{fig1}
\end{figure}

\end{document}